\documentclass[a4paper,11pt]{article}
\usepackage{jcappub} 
\usepackage{lineno}

\title{\boldmath Misner-Sharp Energy and $P$-$V$ Criticality in Quasi-Topological Cosmology}

\author[a]{Yue Chu,}
\author[b]{Shi-Bei Kong,}
\author[a]{Yang Liu,}
\author[c]{Hongsheng Zhang}
\author[a,d]{and Ya-Peng Hu\thanks{Corresponding author}}
\affiliation[a]{College of Physics, Nanjing University of Aeronautics and Astronautics,\\ Nanjing, 211106, China}
\affiliation[b]{School of Science, East China University of Technology, \\Nanchang 330013, Jiangxi, China}
\affiliation[c]{School of Physics and Technology, University of Jinan,\\
 336 West Road of Nan Xinzhuang, Jinan, Shandong 250022, China}
\affiliation[d]{Key Laboratory of Aerospace Information Materials and Physics (NUAA), MIIT, \\Nanjing 211106, China}

\emailAdd{huyp@nuaa.edu.cn}

\abstract{We presented a sound foundation of thermodynamics for a Friedmann-Robertson-Walker (FRW) universe from the first principle in ground-breaking work [Hu et al., JHEP \\12 (2022) 168].
Based on such an approach, we explore the thermodynamics of cosmology in quasi-topology gravity. 
Starting from the unified first law, we first obtain the well-defined Misner-Sharp energy in quasi-topology cosmology.
We demonstrate that the Misner-Sharp energy is equal to $\rho V$ inside the apparent horizon.
Further, the unified first law requires extra terms for generalized force and conjugate generalized position, which are identified as thermodynamic pressure and thermodynamic volume, respectively.
Hence we naturally derive the equation of state of the FRW universe in quasi-topology gravity, and show that it undergoes $P$-$V$ phase transitions. We calculate the critical exponents for the phase transition, which may be beneficial to probe the micro theory of quasi-topology gravity.
}

\begin{document}
\maketitle
\flushbottom

\section{Introduction}
\label{sec:intro}

Black hole thermodynamics is a significant and far-reaching advance in theoretical physics, which bridges the macroscopic theory of gravity and a possible microscopic theory of gravity still in exploration \cite{Wald:1999vt,Witten:2024upt}. Initially formulated within black hole mechanics, black holes were viewed as purely gravitational objects devoid of thermodynamic attributes \cite{Bardeen:1973gs}.
 This paradigm shifted radically with Hawking’s seminal discovery that black holes emit thermal radiation at a temperature \cite{Hawking:1975vcx}.
 Crucially, this temperature matches the  thermodynamic temperature conjectured from classical black hole mechanics, thereby unifying the two frameworks.

The discovery of Hawking radiation poses a fundamental challenge to black hole thermodynamics: as the event horizon shrinks due to quantum evaporation, the Bekenstein-Hawking entropy decreases monotonically in the forward time direction \cite{Almheiri:2020cfm}. 
This apparent violation of the classical second law necessitates the generalized second law \cite{Bekenstein:1974ax}, which implies that the total entropy, the sum of the black hole entropy and the thermal entropy of Hawking radiation, never decreases.
At the same time one arranges the stress energy tensor of the radiations at right hand side of field equation as an effective source. The metric gets changed accordingly.  Such a backreaction of Hawking radiation is not considered in the original form of black hole thermodynamics.
To unify the thermodynamics of spacetime and matters trapped in a finite region, the unified first law is proposed \cite{Hayward:1993wb,Hayward:1997jp,Hayward:1998hb}, which is a reformulation of Einstein's equations and applicable to finite regions of spacetime without the asymptotic information of the whole manifold. This property makes the unified first law much more practicable in realistic celestial processes.

Extending the unified first law to modified theories of gravity remains an open and non-trivial problem. While certain modified gravities, such as the Gauss-Bonnet gravity \cite{Cai:2005ra}, scalar-tensor/Horndeski gravity \cite{Cai:2006rs} and massive gravity \cite{Hu:2015xva}, appear to admit a direct generalization of the unified first law, their implementation often experiences difficulty. For example, in Horndeski theories, the existence of the unified first law depend on
 specific integrability constraints \cite{Cai:2006rs}.
On the other hand, In cosmological contexts, the first law may acquire an additional entropy production term \cite{Cai:2009qf,Zhang:2014goa}, reflecting essentially non-equilibrium process.
Despite these complexities, one observes from the unified first law is that one can obtain the proper form of Misner-Sharp energy \cite{Misner:1964je} in modified gravities, which is consistent to the result derived from the  conserved charge approach  \cite{Cai:2009qf}. The remarkable merit of unified first law is that it can be applied to dynamical spacetime without conceptual problem, for example the FRW universe, since it is just a reformulation of the field equation.

The thermodynamics of the FRW universe is studied by treating the apparent horizon as an analogy of the black hole event horizon\cite{Cai:2006rs}.
Seminal work \cite{Cai:2005ra,Cai:2008gw,Hu:2010tx} reveal that, within semiclassical gravity, the apparent horizon emits Hawking-like radiation,  endowing the FRW spacetime with thermodynamic attributions \cite{Pathria:1996hda,Cai:2005ra,Akbar:2006er,Akbar:2006kj,Cai:2006rs,Gong:2007md,Cai:2008gw,Hu:2010tx}.
The Misner-Sharp energy for a FRW universe naturally appears the unified first law~\cite{Cai:2006rs}. Note that, in the first law of thermodynamics, extra terms for generalized force and conjugate generalized coordinate emerge naturally. Therefore, the generalized force conjugating to volume can be reasonably identified as pressure. Following this theoretical observation, the definition of thermodynamic  pressure in AdS black holes is gracefully solved, where thermodynamic  pressure has been found to relate to the negative cosmological constant in previous studies of $P$-$V$ phase transition \cite{Kubiznak:2012wp}. It should be emphasized that, by using the unified first law, we have further proposed a generic form of thermodynamic  pressure in the scenario of cosmology as well as black holes. That is, after projecting the unified first law on the apparent horizon of the FRW universe, one clarifies the first law of thermodynamics, where the work density plays the role of the thermodynamic pressure.

By using the generic form of thermodynamic  pressure, we independently  obtain the equation of state for a FRW universe which satisfies $P=P(T,V)$, where $T$ is the temperature, and $V=4\pi R_A^3/3$ is the thermodynamic volume of the $4$-dimensional FRW universe \cite{Kong:2021dqd}. Precisely due to the significance of the equation of state (as is
  well-known in van der Waals systems and black hole thermodynamics \cite{Kubiznak:2012wp}), we identify it  within the FRW universe and utilize it to investigate phase transitions.
  Furthermore, we have obtained the equations of state and investigated the $P$-$V$ phase transitions
  for the FRW universe in several theories of gravity
  \cite{Abdusattar:2021wfv,Kong:2021dqd,Abdusattar:2023hlj,Kong:2022xny,Abdusattar:2022bpg},
  and found a few interesting results. For example, in Einstein gravity and brane world scenario, there is no $P$-$V$ phase
  transition \cite{Kong:2021dqd}.
  In contrast,  in scalar-tensor theories of gravity we found $P$-$V$ phase transitions
  and obtained the critical exponents, which satisfy the scaling laws \cite{Hu:2024ldp}.

It is of interest to consider the higher curvature terms for gravitational theories, which typically leads to higher-order equations of motion \cite{Bueno:2016ypa}. Specially, Gauss-Bonnet (GB) gravity still remains second-order equations of motion even when higher curvature
terms are included. A straightforward extension of GB gravity is Lovelock gravity \cite{Lovelock:1971yv}, while the second-order
Lovelock term is known as the GB term. Lovelock gravity also successfully preserves the form of second-order equations of motion, thereby avoiding
instabilities and additional degrees of freedom associated with higher-order differential equations.
{Despite having various advantages, its application is restricted by a minimum number of spacetime dimensions.
In particular, the Lovelock term of order $i$ vanishes identically to $d\leq 2i-1$, and becomes a total derivative that does not contribute to the equations of motion for $d = 2i$.
A total derivative further reduces to a surface term in the action and only contributes to topological
characteristics, so a higher curvature term of this kind is also called a topological term. 

A natural question is whether it is possible to construct a gravitational action that includes cubic or
higher-order curvature interactions, which can contribute to local dynamics in lower-dimensional spacetimes.
This goal has been achieved by Myers et al. in the theory of quasi-topological gravity (QTG)
\cite{Myers:2010ru,Dehghani:2011vu}, which is often used as an insightful toy model for the holographic study on CFT with $d\geq 4$.
Specifically, the cubic curvature term in quasi-topological gravity contributes to the field
equations starting from 5-dimensional spacetime rather than seven.
This extension makes some terms in Lovelock gravity become sensible in lower dimensions.
Research about black hole solutions and holography in QTG has been extensively conducted \cite{Ali:2024rtb,Ali:2023gny,Olamaei:2023csx,Marks:2023ipa,Chen:2022fdi,Naeimipour:2021bgc,Dehghani:2013ldu}, but it has rarely been studied in the context of cosmology. Recently, the thermodynamics of quasi-topological cosmology has been well investigated \cite{Dehghani:2013oba}. Regarding the thermodynamic properties of the apparent horizon in a FRW universe governed by QTG,
the authors investigate its first law of thermodynamics, and derive the general form of Friedmann equation. 

In our paper, we further explore the thermodynamics of quasi-topological cosmology from a new perspective. We first derive the Misner-Sharp energy in quasi-topological cosmology through investigating the unified first law. From this form, we demonstrate that the Misner-Sharp energy within the apparent horizon is indeed $\rho V$ which is just used as an assumption in previous research. In addition, after taking the first law of thermodynamics at the apparent horizon into account, we further make a proper definition of thermodynamic pressure $P$ for the FRW universe in quasi-topological gravity similar to our previous work \cite{Abdusattar:2021wfv,Kong:2021dqd,Abdusattar:2023hlj,Kong:2022xny,Abdusattar:2022bpg}, and hence obtain the equation of state $P=P(T,V)$ in quasi-topological cosmology. Furthermore, we discover that there is an interesting $P$-$V$ criticality in quasi-topological cosmology, where the corresponding critical exponents are as the same as those in the mean field theory.

Our paper is organized as follows.
In Sec.II, we derive the Misner-Sharp energy of quasi-topological cosmology from the unified first law. In Sec.III, we obtain the equation of state $P=P(T,V)$ for the quasi-topological cosmology and use it to study $P$-$V$ criticality.
Sec.V is devoted to conclusion and discussions.

\section{Misner-sharp energy in quasi-topological cosmology}

In this section, we derive the Misner-Sharp energy of a FRW universe in quasi-topological gravity by investigating the unified first law. The Misner-Sharp energy and unified first law are critical for further investigations on the thermodynamics and equation of state $P=P(V,T)$
in quasi-topological cosmology. 

\subsection{A review of quasi-topological cosmology}

A general action of ($n+1$)-dimensional quasi-topological gravity is given by \cite{Dehghani:2011vu}
\begin{equation}
I=I_G+I_m=\int d^{n+1}x\frac{\sqrt{-g}}{16 \pi G_{n+1}} (\mu_1\mathcal{L}_1+\mu_2
\mathcal{L}_{2}+\mu_3 \chi_3+\mu_4
\chi_4)\ +I_m\ , \label{action}
\end{equation}
where $I_G$ and $I_m$ are the action of gravitational and matter field, respectively. 
In Eq. (\ref{action}), $\mu_i$ are four constants, $\mathcal{L}_1=R$ is the Einstein-Hilbert term,
$\mathcal{L}_{2}=R_{abcd}R^{abcd}-4R_{ab}R^{ab}+R^2$ is the Gauss-Bonnet (the second-order Lovelock) term, $\chi_3$
is the curvature-cubed term  \cite{Myers:2010ru}
\begin{eqnarray}
\chi_{3}&=&R_{abcd}R^{bedf}R_e{}^a{}_f{}^c+\frac{1}{(2n-1)(n-3)}\biggl( \frac{3(3n-5)}{8}R_{abcd}R^{abcd}R \nonumber
\\
&&-3(n-1)R_{abcd}R^{abc}{}_e R^{de}+3(n+1)R_{abcd}R^{ac}R^{bd} \nonumber \\
&&+6(n-1)R_{ab}R^{bc}R_c{}^a-\frac{3(3n-1)}{2}R_{ab}R^{ab}R+\frac{3(n+1)}{8}R^3 \biggl),
\end{eqnarray}
and $\chi_4$ is constructed as \cite{Dehghani:2011vu}
\begin{eqnarray}
\chi_{4}&=&c_{1}R_{abcd}R^{cdef}R^{hg}{}_{ef}R_{hg}{}^{ab}+c_{2}R_{abcd}R^{abcd}R_{ef}R^{ef}
+c_{3}RR_{ab}R^{ac}R_{c}{}^{b}+c_{4}(R_{abcd}R^{abcd})^{2}  \nonumber \\
&&+c_{5}R_{ab}R^{ac}R_{cd}R^{db}+c_{6}RR_{abcd}R^{ac}R^{db}+c_{7}R_{abcd}R^{ac}R^{be}R^{d}{}_{e}
+c_{8}R_{abcd}R^{acef}R^{d}{}_{e}R^{d}{}_{f} \nonumber \\
&&+c_{9}R_{abcd}R^{ac}R_{ef}R^{bedf}+c_{10}R^{4}+c_{11}R^{2}R_{abcd}R^{abcd}+c_{12}R^{2}R_{ab}R^{ab} \nonumber
\\
&&+c_{13}R_{abcd}R^{abef}R_{ef}{}^{c}{}_{g}R^{dg}+c_{14}R_{abcd}R^{aecf}R_{gehf}R^{gbhd}\ .
\end{eqnarray}
Here coefficients $c_i$ are specifically chosen to maintain the equations of motion to be second-order, which are
\begin{align}
c_{1}&=-(n-1)(n^7-3n^6-29n^5+170n^4-349n^3+348n^2-180n+36), \nonumber \\
c_{2}&=-4(n-3)\left(2 n^{6}-20 n^{5}+65 n^{4}-81 n^{3}+13 n^{2}+45 n-18\right),  \nonumber \\
c_{3}&=-64 (n-1) \left(3 n^{2}-8 n+3\right)\left(n^{2}-3 n+3\right), \nonumber \\
c_{4}&=-(n^8-6n^7+12n^6-22n^5+114n^4-345n^3+468n^2-270n+54), \nonumber \\
c_{5}&=16 (n-1)\left(10 n^4-51 n^3+93 n^2-72 n+18\right),\nonumber \\
c_{6}&=-32 \left(n-1\right)^{2}\left(n-3\right)^{2}\left(3 n^{2}-8 n+3\right),\nonumber \\
c_{7}&=64 \left(n-2\right)\left(n-1\right)^{2}\left(4 n^{3}-18 n^{2}+27 n-9\right), \nonumber \\
c_{8}&=-96 \left(n-1\right)\left(n-2\right)\left(2 n^{4}-7 n^{3}+4 n^{2}+6 n-3\right), \nonumber \\
c_{9}&=16(n-1)^{3}\left(2 n^{4}-26 n^{3}+93 n^{2}-117 n+36\right), \nonumber \\
c_{10}&=n^{5}-31 n^{4}+168 n^{3}-360 n^{2}+330 n-90,   \nonumber \\
c_{11}&=2(6n^{6}-67 n^{5}+311 n^{4}-742 n^{3}+936 n^{2}-576 n+126),  \nonumber \\
c_{12}&=8(7n^{5}-47 n^{4}+121 n^{3}-141 n^{2}+63 n-9), \nonumber \\
c_{13}&=16n(n-1) (n-2) (n-3) \left(3 n^{2}-8 n+3\right), \nonumber \\
c_{14}&=8(n-1)\left(n^{7}-4 n^{6}-15 n^{5}+122 n^{4}-287 n^{3}+297 n^{2}-126 n+18\right) \nonumber.
\end{align}

For obtaining the Misner-Sharp energy in quasi-topological cosmology, we  first introduce a FRW metric ansatz \cite{Dehghani:2011vu}
\begin{alignat}{1}
d s^2=-N(t)d t^2+a^2(t)\biggl(\frac{d r^2}{1-k r^2}+r^2d\Omega^2_{n-1}\biggr)
\ , \label{metric}
\end{alignat}
where $k$ is the spatial curvature
constant, $d\Omega^2_{n-1}$ represents the line elements of an $(n-1)$-dimensional unit sphere, and $a(t)$ is the scale factor.
In addition, to facilitate variation, we introduce the lapse function $N(t)$, which will be set to 1 after the variation is performed.
 After inserting this metric ansatz into the action (\ref{action})
and integrating by parts, one yields \cite{Myers:2010ru,Dehghani:2011vu}
\begin{equation}\label{a9}
I_{G}=\int d^{n+1}x\frac{n(n-1)}{16\pi
G_{n+1}}\frac{\sqrt{\gamma}a^{n}}{N^{(n+3)/2}a^{8}}\biggl(-N^{4}b_{1}+N^{3}b_{2}\dot{a}^{2}+\frac{1}{3}N^{2}b_{3}\dot{a}^{4}+\frac15Nb_4\dot{a}^6+\frac17b_5\dot{a}^8\biggr)\
,
\end{equation}
where $\gamma$ is the determinant of line element $ds_n^2=d r^2/(1-k r^2)+r^2d\Omega^2_{n-1}$. The coefficients $b_i$ are
\begin{equation*}
\begin{aligned}
&b_{1} =a^{6}k+a^{4}k^{2}\hat{\mu}_{2}l^{2}+a^{2}k^{3}\hat{\mu}_{3}l^{4}+k^{4}\hat{\mu}_{4}l^{6}\ ,
\quad b_{2} =a^6+2a^4k\hat{\mu}_2l^2+3a^2k^2\hat{\mu}_3l^4+4k^3\hat{\mu}_4l^6\ , \\
&b_{3} =a^4\hat{\mu}_2l^2+3a^2k\hat{\mu}_3l^4+6k^2\hat{\mu}_4l^6\ , \quad
b_{4} =a^{2}\hat{\mu}_{3}l^{4}+4k\hat{\mu}_{4}l^{6}, \quad b_{5} =\hat{\mu}_4l^6\ ,
\end{aligned}
\end{equation*}
where $\hat{\mu}_i$ are dimensionless parameters
\begin{equation*}
\begin{aligned}
\hat{\mu}_{1} =&1\ ,\quad \hat{\mu}_2=(n-2)(n-3)\frac{\mu_2}{l^2}\ , \quad \hat{\mu}_3
=\frac{(n-2)(n-5)(3n^2-9n+4)}{8(2n-1)}\frac{\mu_3}{l^4}\ ,
\nonumber \\
\hat{\mu}_{4} =&n(n-1)(n-2)^{2}(n-3)(n-7)(n^{5}-15n^{4}+72n^{3}-156n^{2}+150n-42)\frac{\mu_{4}}{l^6}\ .
\end{aligned}
\end{equation*}
In our paper, we focus on the lowest dimensional case in quasi-topological
gravity for simplicity, i.e. $5$-dimensional spacetime with $n=4$ since some interesting results have been shown in previous work. For example, one observe the first instances of both reentrant phase transitions and thermodynamic singularities in five dimensions of AdS quasi-topological black hole solutions \cite{Hennigar:2015esa}.

\subsection{The Misner-Sharp energy}
In deriving the Misner-Sharp energy, the $(t,t)$ and $(r,r)$ components of gravitational field equation play  crucial roles. Therefore, 
varying the action (\ref{action}) with respect to $g_{00}$ and $g_{11}$, we obtain the corresponding components of gravitational tensor $\mathcal{G}^{\mu\nu}$ in quasi-topological cosmology as
\begin{eqnarray}
\mathcal{G}^{tt}=\frac{6}{ a^8}(k+\dot{a}^{2})(a^6+l^2
a^4\hat{\mu}_2(k+\dot{a}^{2})+l^4a^2\hat{\mu}_3(k+\dot{a}^{2})^2+l^6\hat{\mu}_4(k+\dot{a}^{2})^3,\label{rho}
\end{eqnarray}
\begin{eqnarray}\label{p}
\mathcal{G}^r{}^r=
&-&\frac{3(1-kr^2)}{ a^{10}}[(a^6(k+\dot{a}^{2})-l^4
a^2\hat{\mu}_3(k+\dot{a}^{2})^3-2l^6\hat{\mu}_4(k+\dot{a}^{2})^4+a^7\ddot{a} \nonumber\\ 
&+&2l^2a^5\hat{\mu}_2(k+\dot{a}^{2})\ddot{a}+3l^4a^3\hat{\mu}_3(k+\dot{a}^{2})^2\ddot{a}+4l^6a\hat{\mu}_4(k+\dot{a}^{2})^2\ddot{a}] .
\end{eqnarray}
where $\mathcal{G}^{\mu\nu}$ satisfies the equation of gravitational motion $\mathcal{G}^{\mu\nu}=8\pi G_5 T^{\mu\nu}$ in quasi-topological gravity. 
Note that due to the complexity of the action in Eq.(\ref{action}), the explicit form of $\mathcal{G}^{\mu\nu}$ is not provided. However, by adopting the metric ansatz Eq.(\ref{metric}) and applying the variation method above, we explicit the components of $\mathcal{G}^{\mu\nu}$. 
Further, by considering the matter field as an ideal fluid, we obtain
\begin{equation}\label{FE}
\sum_{i=1}^4\hat{\mu}_il^{2i-2}\biggl(H^2+\frac k{a^2}\biggr)^i=\frac{4\pi G_{5}}{3}\rho\ ,
\end{equation}
and 

\begin{eqnarray}\label{p}
-\sum_{i=1}^4i\hat{\mu}_il^{2i-2}\biggl(H^2+\frac k{a^2}\biggr)^{i-1}\frac{\ddot{a}}{a}+
\sum_{\substack{i=1 \\ i \neq 2}}^4 (i-2)\hat{\mu}_il^{2i-2}\biggl(H^2+\frac k{a^2}\biggr)^i
=\frac{8\pi G_{5}}{3}p\ ,
\end{eqnarray}
where $H\equiv\dot{a}(t)/a(t)$ is the Hubble parameter, $\rho$ and $p$ 
represent the energy density and pressure of a perfect fluid, respectively.
 From Eqs.(\ref{FE}) and (\ref{p}) , we can find that 
 these two equations clearly correspond to the modified Friedmann equations in quasi-topological cosmology
\cite{Dehghani:2011vu}, and reduce to the Einstein gravity case with $\hat{\mu}_2, \hat{\mu}_3$ and $\hat{\mu}_4=0$. 
Furthermore, it can be verified that the energy density $\rho$ and pressure $p$ from Eqs.(\ref{FE}) and (\ref{p}) indeed satisfy the continuity equation of matter field, $\dot{\rho}+4H(\rho+p)=0$.

On the other hand, if the generalized Misner-Sharp energy in modified gravity exists, it satisfies  \cite{Hayward:1993wb,Zhang:2014goa}
\begin{equation}\label{ufl1}
dE_\mathrm{MS}=A'\Psi_adx^a+Wd V'\ ,
\end{equation}
where $A'=n \Omega_n R^{n-1}$ is the area of the $n$-dimensional sphere with radius $R$, and $V'=\Omega_n R^{n}$ is its volume. The work density is
$W=-h^{ab}T_{ab}/2=(\rho-p)/2$, while the energy-supply vector is defined as $\Psi_{a}=T_{a}{
}^{b}\partial_{b}R+W\partial_{a}R$, where $T_{ab}$ is the energy-momentum tensor $T_{\mu\nu}$ projecting onto 2-dimensional space-time normal to
the sphere. Therefore, from the right-hand side of Eq.(\ref{ufl1}), it can be explicitly
expressed as \cite{Hu:2015xva}
\begin{equation}\label{ufl}
A'\Psi_adx^a+WdV'=\tilde A(t,r)dt+\tilde B(t,r)dr\ ,
\end{equation}
where
\begin{equation}
\tilde A(t,r)=A'(T_t{}^r R,_r-T_{r}{}^rR,_t)\ ,~\tilde B(t,r)=A'(T_t{}^r R,_t-T_{t}{}^tR,_r)\ ,
\end{equation}
with $R=a(t)r$.
After considering $n=4$, i.e. $A'_4=2\pi^2 R^3$ and $V'_4=\pi^2 R^4/2$ and the  field equations Eqs.(\ref{FE}) and (\ref{p}) with $T_t^{~r}=T_r^{~t}=0$, we precisely write $\tilde A(t,r)$ and $\tilde B(t,r)$ as
\begin{eqnarray}
\tilde A(t,r)&=&\frac{1}{4 G_5 a^5}3\pi r^4 \dot{a}(a^6(k+\dot{a}^{2})-l^4
a^2\hat{\mu}_3(k+\dot{a}^{2})^3-2l^6\hat{\mu}_4(k+\dot{a}^{2})^4+a^7\ddot{a} \nonumber\\
&&+2l^2a^5\hat{\mu}_2(k+\dot{a}^{2})\ddot{a}+3l^4a^3\hat{\mu}_3(k+\dot{a}^{2})^2\ddot{a}+4l^6a\hat{\mu}_4(k+\dot{a}^{2})^2\ddot{a}
), \nonumber\\
\tilde B(t,r)&=&\frac{1}{2 G_5 a^4}3\pi r^3(k+\dot{a}^{2})[a^6+l^2 a^4\hat{\mu}_2(k+\dot{a}^{2})+l^4
a^2\hat{\mu}_3(k+\dot{a}^{2})^2+l^6 \hat{\mu}_4(k+\dot{a}^{2})^3].\nonumber
\end{eqnarray}
From this, we can easily verify that the following
integral condition
\begin{equation}
\frac{\partial \tilde A(t,r)}{\partial r}=\frac{\partial \tilde B(t,r)}{\partial t}\ ,
\end{equation}
 is automatically satisfied.
Therefore, the generalized Misner-Sharp energy in quasi-topological gravity is rigorously derived, yielding:
\begin{eqnarray}
E_\mathrm{MS}
&=&\int \tilde B(t,r)dr+\int\left[\tilde A(t,r)-\frac\partial{\partial t}\int \tilde B(t,r)dr\right]dt \nonumber\\ 
&=&\frac{3\pi r^4}{8 G_5 a^4}(k+\dot{a}^{2})[a^6+l^2 a^4\hat{\mu}_2(k+\dot{a}^{2})+l^4
a^2\hat{\mu}_3(k+\dot{a}^{2})^2+l^6 \hat{\mu}_4(k+\dot{a}^{2})^3] \nonumber\\ 
&=&\frac{3 \pi R^4}{8 R_A^2 G_5}\left(1+\hat{\mu}_2\frac{l^2}{R_A^2}+\hat{\mu}_3\frac{l^4}{R_A^4}+\hat{\mu}_4\frac{l^6}{R_A^6}\right)\ ,
\end{eqnarray}
where $R_A=1/\sqrt{\frac{\dot{a}^2}{a^2}+\frac{k}{a^2}}$ is the radius of the apparent horizon for a FRW universe. 
A notable aspect of this work is that the generalized Misner-Sharp energy, derived here, is applicable to a sphere of arbitrary radius 
$R$ within the context of quasi-topological cosmology, an advancement that has not been previously established.
Furthermore, we demonstrate that the Misner-Sharp energy inside the apparent horizon 
$R_A$ is exactly $\rho V$, a result often assumed in prior studies. This provides a more rigorous foundation for this commonly used assumption.

\section{The Equation of state and $P$-$V$ criticality in quasi-topological cosmology}
In the previous section, we derived the generalized Misner-Sharp energy in quasi-topological cosmology, and as a result, Eq. (\ref{ufl1}) represents the generalized unified first law.
Notably, by projecting this generalized unified first law onto the apparent horizon, one recovers the first law of thermodynamics, as demonstrated in \cite{Dehghani:2013oba}. 
Building on this foundational result, the current section aims to further explore the thermodynamic properties within quasi-topological cosmology. 
Specifically, we mainly focus on the equation of state $P=P(V,T)$ and $P$-$V$ criticality in quasi-topological cosmology, which are crucial aspects of thermodynamics that have not yet been investigated.

\subsection{The equation of state }
A pivotal requirement for deriving the equation of state of a gravitational spacetime lies in a well-founded definition of thermodynamic pressure. Drawing from our earlier work \cite{Kong:2021dqd}, we formulate this definition through the generalized unified first law or the first law of thermodynamics within quasi-topological cosmology
, which is
\begin{eqnarray}
dE=-T_h dS_h+W d V \ .
\end{eqnarray}
Here $E=E_\mathrm{MS}|_{R=R_A}=\rho  V$ is the Minser-Sharp energy inside the apparent horizon $R_A$ where  $V=V'_4|_{R=R_A}=\pi^2 R_A^4/2$,  $T_h=1/2\pi R_A(1-\dot{R}_A/2)$ is the Hawking temperature of a FRW universe. In addition, $W$ is the above work density, while the entropy $S_h$ is
\begin{eqnarray}
S_{h}=\frac{A}{4G_{5}}\biggl(\hat{\mu}_{1}+\frac{6\hat{\mu}_{2}l^{2}}{R_{A}^{2}}-\frac{9\hat{\mu}_{3}l^{4}}{R_{A}^{4}}-\frac{4\hat{\mu}_{4}l^{6}}{R_{A}^{6}}\biggl)\ ,
\end{eqnarray}
which has the same form as the black hole entropy in quasi-topological gravity,    
      except that the event horizon radius of the black hole is replaced by the radius of the apparent horizon.

Comparing with the standard first law of thermodynamics $d U=Td S-Pd V$, one can read out the internal energy and pressure $P$ of the FRW universe in quasi-topological
theory as \cite{Kong:2021dqd}
\begin{alignat}{1}
U\equiv -E, \quad P\equiv W. \quad
\end{alignat}
It is important to note that this definition of 
$P$ is derived from first principles, and thus, it can be appropriately regarded as the thermodynamic pressure for FRW spacetime within the context of QTG.
Therefore, the equation of state for the thermodynamic system inside the apparent horizon of a FRW universe is
\begin{eqnarray}\label{eoso}
P(T,R_A)=\frac{3T}{4R_A}+\frac{3}{8\pi R_A^2}+\frac{3T\alpha}{2R_A^3}+\frac{9T\beta}{4R_A^5}-\frac{3\beta}{8\pi R_A^6}
+\frac{3T\gamma}{R_A^7}-\frac{3\gamma}{4\pi R_A^8}\ ,
\end{eqnarray}
where $\alpha=\hat{\mu}_2l^{2}, \beta=\hat{\mu}_3l^{4}$ and $\gamma=\hat{\mu}_4l^{6}$ are three new parameters defined for convenience. 
Obviously, if these three parameters are all zero, the corresponding thermodynamic equation of state of the $5$-dimensional FRW universe in Einstein gravity is obtained 
\begin{eqnarray}\label{eoso1}
P(T,R_A)=\frac{3T}{4R_A}+\frac{3}{8\pi R_A^2} ,
\end{eqnarray}
which is similar with the result of the $4$-dimensional FRW universe in Einstein gravity \cite{Kong:2021dqd}.

\subsection{The $P$-$V$ phase transition and criticality}
From the equation of state in Eq.(\ref{eoso}), we will study the phase transition and criticality in this subsection. The two conditions of critical points are
\begin{eqnarray}\label{towc}
{\left(\frac{\partial P}{\partial R_A}\right)}_T=0, \quad {\left(\frac{{\partial}^2 P}{\partial R_A^2}\right)}_T=0\ .
\end{eqnarray}
After using the equation of state, those two conditions are further presented
\begin{eqnarray}
\pi R_A^{7}T+R_A^{6}+6\pi R_A^{5}T\alpha+15\pi R_A^{3}T\beta-3R_A^2\beta+28\pi R_A T\gamma-8\gamma=0\ ,
\nonumber \\ 
2\pi R_A^{7}T+3R_A^{6}+24\pi R_A^{5}T\alpha-21R_A^{2}\beta +90\pi R_A^3 T\beta+224\pi R_A T \gamma-72\gamma=0\ .
\end{eqnarray}
Thus, we conclude that the critical radius of the apparent horizon satisfies the following condition
\begin{eqnarray}\label{rc}
&&-R_c^{12}+6R_c^{10}\alpha+60R_c^{8}\beta+2R_c^{6}(27\alpha \beta+98\gamma) 
+15R_c^{4}(3\beta^2+16\alpha\gamma)\nonumber\\
&&+276R_c^{2}\beta \gamma+224 \gamma^2=0\ ,
\end{eqnarray}
where $R_c$ should not be zero from the equation of state. From which, we can see that Eq.(\ref{rc}) contains higher-order terms of $R_c$, thus it is usually difficult to have analytical solutions. For convenience in later investigation, we express the critical pressure and temperature in terms of $R_c$ as
\begin{eqnarray}
P_c&=&\frac{3(-R_c^{12}+2R_c^{10}\alpha+14R_c^{8}\beta+R_c^{6}(6\alpha \beta+34 \gamma)+R_c^{4}(3\beta^2+20\alpha \gamma)+14R_c^{2}\beta \gamma+8\gamma^2)}{8\pi R_c^{8}(R_c^{6}+6R_c^{4}\alpha+15R_c^{2}\beta+28\gamma)}\ ,\nonumber \\
T_c&=&-\frac{R_c^{6}-3R_c^{2}\beta-8\gamma}{\pi R_c^{7}+6\pi R_c^{5}\alpha+15\pi R_c^{3}\beta+28\pi R_c\gamma}\ . 
\end{eqnarray}

Note that the existence of critical radius in Eq.(\ref{rc}) depends on the three parameters $\alpha$, $\beta$ and $\gamma$. Obviously, if these three parameters are all zero, there is no critical radius, which corresponds to the absence of $P$-$V$ phase transition. This situation is also similar to the case of the $4$-dimensional FRW universe in Einstein gravity. Therefore, the existence of critical radius and its corresponding properties like $P$-$V$ phase transition is an interesting open issue. For the sake of simplicity, we focus primarily on the case with $\alpha,\gamma=0$, thus the corresponding equation of critical radius is simplified as
\begin{eqnarray}\label{rc1}
R_c^{8}-60R_c^{4}\beta
-45\beta^2=0\ .
\end{eqnarray}
Significantly, a physically relevant critical radius with a positive value is observed when $\beta$ is negative. Moreover, the corresponding critical temperature and pressure are demonstrated to be physically meaningful, as both parameters maintain positive values. The exact analytical expressions for these critical quantities are
\begin{eqnarray}
R_c&=&(3\times(\sqrt{105}-10))^\frac{1}{4}(-\beta )^{1/4},\quad
T_c=\frac{(\sqrt{{35/3}}-{10/3})^{3/4}(11\sqrt{3}+3\sqrt{35})}{20\pi(-\beta )^{1/4}
},\quad \nonumber \\
P_c&=&\frac{95\sqrt{3}-27\sqrt{35}}{120(\sqrt{105}-10)^\frac{3}{2}\pi \sqrt{-\beta}}\ ,
\end{eqnarray}
which indicates
the existence of a $P$-$V$ phase transition in this case.
In Fig.\ref{fig1}, we have clearly illustrated the case of the phase transition, where $\beta=-1$ is chosen for convenience. The corresponding critical point is $R_c=0.928,~T_c=0.090$, and$~P_c=0.104$. 


\begin{figure}[h]
  \centering
  \includegraphics[width=0.7\textwidth]{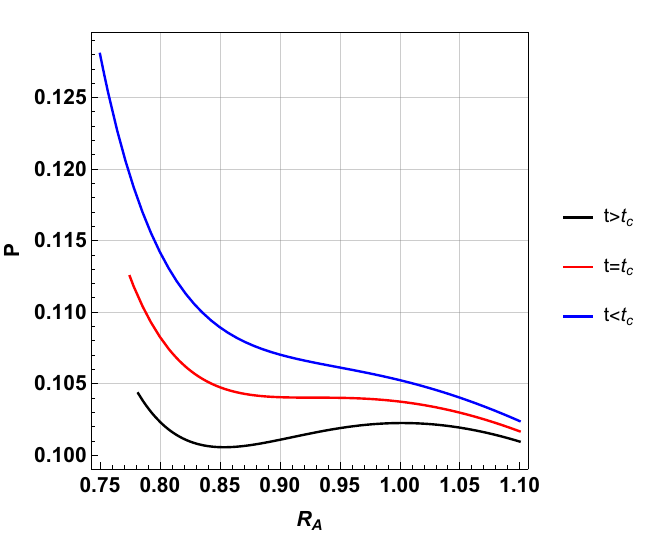}
  \caption{$P$-$R_A$ diagram illustrating the phase transition in quasi-topological cosmology. The diagram presents three isotherms corresponding to \(T > T_c \) (black), \(T = T_c \) (red), and \(T < T_c \) (blue). It is evident that its behavior near the critical point is the exact opposite of that of typical Van der Waals fluids, where phase coexistence occurs at temperatures below the critical temperature.
 }\label{fig1}
\end{figure}

It should be emphasized that other parameter choices also allow for the existence of a physical critical point.
For example, when $\alpha,\beta=0$ and $\gamma=-1$, the corresponding critical point is $R_c=1.102$, $T_c=0.106$ and $P_c=0.120$, while for $\alpha=1$,  $\beta=1$ and $\gamma=-1$, the critical point is $R_c=0.736$, $T_c=0.157$ and $P_c=0.598$. Nevertheless, 
we observe that these cases exhibit similar $P$-$V$ phase transition behavior and, therefore, do not present them in detail.
 In addition, there is one interesting difference from ordinary thermodynamic system. That is, 
the coexistence curve occurs above instead of below the critical temperature.
This phenomenon also appears in previous results of the criticality of the FRW universe, which is 
a typical feature of the FRW universe.

Similarly with the Van der Waals case, we further find that the phase transition at the critical point is second-order, and hence we can further investigate the $P$-$V$ criticality and its critical exponents in quasi-topological cosmology. The four critical exponents are often
defined as
\begin{eqnarray}
C_{V}&=&T{\left(\frac{\partial S}{\partial T}\right)}_{V}\sim \left| t \right|^{-\tilde{\alpha}} ,
 \nonumber \\
\eta&=&\rho_{l}-\rho_{s}\sim v_{l}-v_{s}\sim|t|^{\tilde{\beta}}  ,
 \nonumber\\
\kappa_{T}&=&-\frac{1}{V}{\left(\frac{\partial V}{\partial P}\right)}_{T}\sim\left| t
\right|^{-\tilde{\gamma}},
 \nonumber\\
| P&-&P_{c}| \sim \left| \rho-\rho_{c} \right|^{\delta},
\end{eqnarray}
where `$l$' and `$s$' represents `large' and `small' phases, respectively. $C_V$ is the heat capacity at constant volume, $\eta$ is the order parameter, 
$\kappa_{T}$ is the isothermal compressibility, and the last one $|P-P_{c}|$ shows the critical behavior of the thermodynamic pressure. 

Following previous work \cite{Kong:2021dqd}, there is a simple method to calculate these four critical exponents. First, one can expand the equation of state around the critical point as 
\begin{eqnarray}
\tilde{p}=a_{10}t+a_{11}tv+a_{03}v^3+o(tv^2,v^4)\ , \label{expansion}
\end{eqnarray}
where $t=T/T_c-1, \tilde{p}=P/P_c-1, v=V/V_c-1$, and the corresponding coefficients are
\begin{eqnarray}
a_{10}&=&\left(\frac{\partial \tilde{p}}{\partial t}\right)_{c}=\frac{T_c}{P_c}\left(\frac{\partial P}{\partial
T}\right)_c,
\\
a_{11}&=&\left(\frac{\partial^2 \tilde{p}}{\partial t\partial v}\right)_{c}
=\frac{R_c T_c}{3P_c}\left(\frac{\partial^2 P}{\partial T\partial R_A}\right)_c,
\\
a_{03}&=&\frac{1}{3!}\left(\frac{\partial^3 \tilde{p}}{\partial v^3}\right)_{c}
=\frac{R_c^3}{162P_c}\left(\frac{\partial^3 P}{\partial R_A^3}\right)_c\ ,
\end{eqnarray}
where the two defining conditions Eq.(\ref{towc}) of the critical point
have been considered, and hence $a_{01}$ and $a_{02}$ vanish. Therefore, if $a_{11}$ and $a_{03}$ are both nonzero, it can be proved that the four critical exponents are
$\tilde{\alpha}=1,~\eta=1/2,~\kappa=1,~ \delta=3$, which are just results in the mean field theory.

\section{Conclusion and Discussions}
In this study, we derive the Misner-Sharp energy and investigate the $P$-$V$ phase transition and critical phenomena in quasi-topological cosmology. We prove that the Msiner-Sharp energy is well-defined in cosmology for quasi-topological gravity, which is obtained through an integration method based on the unified first law in quasi-topological cosmology.  We demonstrate that the Misner-Sharp energy within the apparent horizon of a FRW universe in QTG precisely equals $\rho V$. By applying the first law of thermodynamics, we identify the thermodynamic pressure in quasi-topological cosmology and subsequently derive its equation of state, $P=P(V,T)$. Notably, we observe that the equation of state can exhibit $P$-$V$ criticality under specific conditions, indicating the presence of a second-order phase transition characterized by a critical point. The critical exponents associated with this criticality are found to be consistent with those of mean field theory.

Several interesting remarks deserve further discussion. First, in contrast to the Van der Waals system, the $P$-$V$ phase transition in quasi-topological cosmology manifests above the critical temperature. This distinctive feature raises intriguing questions regarding the underlying physical mechanisms, which remain an open issue for future theoretical and observational investigations. Second, while the equation of state for conventional systems can typically be derived from microscopic properties using statistical physics, the microscopic theory of gravity remains elusive. Consequently, the  equation of state we derived for quasi-topological cosmology may provide valuable insights into the microscopic structure of gravity or quasi-topological cosmology itself. Finally, recent studies suggest that scaling laws in gravitational systems may be violated under specific conditions, particularly when $a_{11}=0$ \cite{Hu:2024ldp}. This prompts an intriguing question that whether similar violations of scaling laws occur in quasi-topological cosmology. Exploring this possibility represents a compelling direction for future research.


\acknowledgments

We are grateful to Dr.Yihao Yin and Zhi Wang for helpful discussions. Ya-Peng Hu is supported by
National Natural
Science Foundation of China (NSFC) under grant Nos.12175105. Hongsheng Zhang is supported by National Natural Science Foundation of China Grants Nos.12275106 and 12235019.
Shi-Bei Kong is supported by National Natural Science Foundation of China (NSFC) under grant No.12465011 and East China University of Technology (ECUT) under grant DHBK2023002.





\end{document}